\documentclass{article}
\usepackage{graphicx} 
\usepackage[T1]{fontenc}
\usepackage{amsmath}
\usepackage[english]{babel}
\usepackage{array,tabularx}
\usepackage{booktabs} 
\usepackage{makeidx}  
\usepackage{xparse}
\usepackage{moredefs}
\usepackage{lips}
\usepackage{epstopdf} 
\usepackage{color}
\usepackage{csquotes}
\usepackage{xcolor}
\usepackage{times}
\usepackage[hidelinks]{hyperref}
\usepackage[nolist,nohyperlinks]{acronym}
\usepackage{comment}
\usepackage{paralist}
\usepackage{cleveref}
\usepackage{wrapfig}
\usepackage{multirow}
\usepackage{microtype}
\usepackage{listings}
\usepackage[super]{nth}
\usepackage{natbib}
\usepackage{subfig}
\usepackage{float}

\usepackage{authblk}

\title{To the Moon: Analyzing Collective Trading Events on the Wings of Sentiment Analysis}

\author[1]{Tim Matthies\footnote{tmatthie@physnet.uni-hamburg.de}}
\author[2]{Thomas Löhden}
\author[2]{Stephan Leible}
\author[2]{Jun-Patrick Raabe}
\affil[1]{Universit\"at Hamburg, Department of Physics, 20355 Hamburg, Germany}
\affil[2]{Universit\"at Hamburg, Department of Informatics, 20355 Hamburg, Germany}
\date{}                     
\begin{document}

\maketitle
\setcounter{footnote}{0}

\begin{abstract}
This research investigates the growing trend of retail investors participating in certain stocks by organizing themselves on social media platforms, particularly Reddit. Previous studies have highlighted a notable association between Reddit activity and the volatility of affected stocks. This study seeks to expand the analysis to Twitter, which is among the most impactful social media platforms. To achieve this, we collected relevant tweets and analyzed their sentiment to explore the correlation between Twitter activity, sentiment, and stock volatility. The results reveal a significant relationship between Twitter activity and stock volatility but a weak link between tweet sentiment and stock performance. In general, Twitter activity and sentiment appear to play a less critical role in these events than Reddit activity. These findings offer new theoretical insights into the impact of social media platforms on stock market dynamics, and they may practically assist investors and regulators in comprehending these phenomena better.

\end{abstract}

\section{Introduction}
\label{sec:introduction}

 In 2021, the stock market witnessed several notable events that appear to be linked to social media. Among these, the GameStop short squeeze gained the most media attention. During this time, retail investors collectively purchased large amounts of GameStop stock, originating from the social media platform Reddit. Despite the lack of exceptional earnings or news, GameStop's stock price increased by over 20 times while retail investors coordinated and shared their buying activity on various social media platforms, as described by \cite{chohan2021counter}. This period was characterized by an anti-establishment attitude, community, and coordinated mobilization of capital, which we find particularly intriguing.

Given the critical role of sentiment in motivating investment decisions during these events, including this factor in our analysis of their impact is essential. Although a direct link between the stock price increase and social media attention may exist, we must further explore the exact nature of this link. \cite{lyocsa2022yolo} made a good start by showing a significant relationship between Reddit activity and stock price volatility. They used the term "YOLO-Trading" to describe the investment decisions made during these events, often driven by a fear of missing out without fully considering the risks and consequences.

The observed decentralized, yet collective behavior strongly resembles herd behavior in financial markets, which we find fascinating. The popular meme "To the moon" expressed the hope that the stock would continue to rise, and we coined the "MOON" variable as a reference to this sentiment. Although the movement originated on Reddit, it is unclear whether Reddit remained the center of influence on the stock market during the development of this event. Therefore, our study aims to extend this analysis by focusing on Twitter, which has a broader reach and a more prominent media presence than Reddit.

We believe that Twitter as a social media platform which is particularly present in the media is important in this case. The nature of these events requires a large number of people to participate, and Twitter is an excellent platform for users to broadcast their posts to the world. The importance of Twitter data for financial markets has been studied by \cite{sul2014trading}, \cite{tafti2016real}, and \cite{abendschein2022cashtag}. Therefore, we analyzed the general sentiment on Twitter during these events by gathering 2 million tweets from October 2020 to June 2021 using the Twitter API  and the sentiment analysis tool VADER (Valence Aware Dictionary for sEntiment Reasoning) (\cite{hutto2014vader}) to calculate the sentiment of these tweets. We collected tweets from four stocks: GameStop, AMC, Nokia, and BlackBerry. All of them experienced unusually high price fluctuation in the time frame mentioned above.

We suggest a significant relationship between a stock's price fluctuations and the activity or sentiment on Twitter regarding that stock experiencing such an event. Therefore, we aim to answer the following research question: 
What is the relationship between the sentiment analysis of Twitter data and stock price variation? 
The remainder of this paper is structured as follows: In chapter 2, we discuss relevant literature for our study, particularly the work of \cite{lyocsa2022yolo}. In chapter 3, we describe our research method. In chapter 4, we state our results. In chapter 5, we discuss their impact on relevant actors such as brokers, stock traders, and market regulators, and in chapter 6, we present our concluding remarks.

\section{Related Work}
\label{sec:relatedwork}

\subsection{Sentiment analysis for financial interest}

Recent studies have shown that sentiment analysis can be an effective tool for predicting stock market movements and prices. Professional market participants, as well as general sentiment expressed through social media platforms like Twitter, have been found to predict future stock market behavior strongly (\cite{chen2014wisdom}; \cite{zheludev2014can}; \cite{subramanian2022sentiment}; \cite{bollen2011twitter}).
However, the volume of social media posts can limit the usefulness of sentiment analysis, as there may not be enough data to make a broad analysis. In our study, we address this problem by focusing on social media-driven events, which generate an abundance of tweets. We use Twitter sentiment analysis to explore the relationship between Twitter sentiment and the stock market, a topic that has been extensively researched (\cite{souza2015twitter}).
To conduct our sentiment analysis on a large scale, we used VADER, which has been shown to be especially well-suited for analyzing the sentiment of Twitter data (\cite{pano2020complete}; \cite{elbagir2019twitter}). VADER outperforms other commonly used sentiment analyzers, particularly in the social media domain (\cite{bonta2019comprehensive}). The sentiment of a tweet can be used to draw conclusions about volatility. 

\subsection{Volatility}
Financial risks are often measured using volatility, the standard deviation of a stock's returns over a period. If the volatility is predictable, it can be a valuable tool for risk management, as demonstrated by \cite{christoffersen2000relevant}. However, the predictability of volatility decreases as the forecast horizon increases. According to \cite{dumas2009equilibrium}, publicly available information can be used effectively to improve the prediction of stock price volatility. \cite{behrendt2018twitter} say that the intraday volatility of individual stocks can be affected by Twitter sentiment. \cite{audrino2020impact} also discovered that sentiment and attention variables could have significant predictive power for future volatility.

\subsection{Reddit as the primary source}
The primary source of information for YOLO-Events is the website Reddit and its subreddit "r/WallStreetBets." \cite{lyocsa2022yolo} analyzed the correlation between Reddit activity and market price variation of stocks experiencing YOLO-Events. The study found that increased activity on the subreddit "r/WallStreetBets" led to increased stock price volatility for a stock undergoing a YOLO-Event, as measured by the volume of comments indicating an intensifying discussion. \cite{long2023just} extended this analysis by exploring the sentiment of the Reddit discussion about GameStop. They found that the links between the discussion were stronger during bullish periods and weaker during bearish periods.

As we proceed with this research, our analysis will focus on the sentiment expressed on Twitter, which is one of the most influential social media platforms. The relationship between Twitter sentiment and stock price volatility will be explored and compared with previous studies.

\section{Research Method}
\label{sec:researchmethod}
\subsection{Twitter data and sentiment analysis}
For our research, it is necessary to gather a large number of tweets from the period surrounding the GameStop event in order to study overall sentiment on Twitter. To access the Twitter API V2, we utilize the Python library Tweepy \cite{TweepyDocumentation}. We collect tweets from October 2020 to June 2021, with a maximum of 100 tweets every 15 minutes, resulting in a potential total of 2 million tweets. However, not every 15-minute interval contains 100 tweets. Additionally, we filter out tweets with images, videos, or retweets to ensure that we include only the most relevant information. The final number of tweets for each stock can be found in Table 1.
\cref{tab:TweetAmount}. 

We use the following queries for each stock: \enquote{GameStop OR GME,} \enquote{AMC,} \enquote{Blackberry OR \$BB,} and \enquote{Nokia OR NOK.} For BlackBerry, we use the cashtag \enquote{\$BB} instead of the stock symbol \enquote{BB} to filter out irrelevant tweets. Cashtags have been studied as a predictive tool in previous research (\cite{cresci2019cashtag}; \cite{bujari2017cashusing}; \cite{hentschel2014cashfollow}). We sort the tweets by relevancy to obtain the most relevant tweets for each 15-minute interval. After collecting the tweets, we conduct sentiment analysis using the VADER by \cite{hutto2014vader}, exclusively using the compound score. We calculate the average sentiment for each trading day and combine sentiment data for non-trading days with the previous trading day.

\bgroup
\begin{table}[]
\centering
\begin{tabular}{|c|c|c|c|c|}
\hline
                 & GameStop & AMC    & Nokia &  BlackBerry \\ \hline
Number of Tweets & $746\,560$   & $680\,872$ & $294\,608$ & $209\,690$\\ \hline
\end{tabular}
\caption{Amount of tweets collected that were tweeted from October 2020 to June 2021.}
\label{tab:TweetAmount}
\end{table}
\egroup

After collecting the tweets, we proceeded with the sentiment analysis. Using VADER's (\cite{hutto2014vader}) compound score, we calculate the average of the sentiments for each trading day. All the sentiment data for non-trading days will be combined with the previous trading day.

\subsection{Volatility data}
\label{sec:VolCal}
In order to investigate the price variations, we proceed similarly to \cite{lyocsa2022yolo}. The same approximation for the volatility is used, which consists of three averages of different volatility estimations. This approximation uses only daily data, which is an advantage over the conventional volatility calculations that require intraday data. The daily data can be obtained by different services. We use \cite{YFinance} for retrieving the daily stock data. Our daily data consists of the opening price $O_t$, the price high $H_t$, the price low $L_t$, and the closing price $C_t$ for a given trading day $t$. With this, we define the following four quantities:
\begin{align}
    h_t&=\ln{\frac{H_t}{O_t}},& l_t&=\ln{\frac{L_t}{O_t}}, \nonumber \\
    c_t&=\ln{\frac{C_t}{O_t}},& J_t&=\left( \ln{\frac{O_t}{C_{t-1}}} \right)^2.
\end{align}
$h_t$, $l_t$ \& $c_t$ are the logarithm of $H_t$, $L_t$ \& $C_t$ in relation to the opening price $O_t$. The variable $J_t$ is overnight price variation. We will now explain the three estimators: 

\begin{align}
    P_t&=\frac{\left(h_t - l_t \right)^2}{4\cdot \ln{2}},\\
    G_t&=0.511 \left(h_t - l_t \right)^2  -0.383 \left( c_t \right)^2 -0.019 \left( c_t\left( h_t + l_t \right) - 2 h_t  l_t \right),\\
    R_t&= h_t \left(h_t - c_t \right) - l_t \left(l_t - c_t \right).
\end{align}

$P_t$ is the estimator of \cite{parkinson1980extreme}, $G_t$ is the estimator of \cite{garman1980estimation}, and $R_t$ is the estimator of \cite{rogers1991estimating}. Finally, we combine these, adjust for the overnight price variation $J_t$, and take the logarithm to approximate the logarithmic volatility: 
\begin{equation}
    V_t = \ln \left( 100^2 \cdot 252 \left(\frac{1}{3} \left( P_t+G_t+R_t \right) \right) \right)
\end{equation}

\subsection{Regression models}
\label{sec:RegMethod}
We investigate different models to predict the next trading day's volatility using only the previous trading day's data. The models \textbf{M1}, \textbf{M2}, \textbf{M3}, \& \textbf{M4} are identical to that created by \cite{lyocsa2022yolo}. They proposed two parameters: $\text{YOLO}_{1,t}$ and $\text{YOLO}_{2,t}$. $\text{YOLO}_{1,t}$ corresponds to the Reddit activity for a given stock, and $\text{YOLO}_{2,t}$ captures a more general activity of the specific stock using Google Trends. The models are linear regression models where the parameters $\beta_i$ are obtained by Ordinary Least Squares (OLS). The four models are:

\begin{align}
    \textbf{M1: } V_t&=\beta_0+\beta_1 V_{t-1}+\beta_2 M_{t-1}+\beta_3 V I X_{t-1}+\epsilon_t,\\
    \textbf{M2: } V_t&=\textbf{M1}_{t-1}+\beta_4 \text{YOLO}_{1,t-1}+\epsilon_t,\\
    \textbf{M3: } V_t&=\textbf{M1}_{t-1}+\beta_5\text{YOLO}_{2,t-1}+\epsilon_t,\\
    \textbf{M4: } V_t&=\textbf{M1}_{t-1}+\beta_4 \text{YOLO}_{1,t-1}+\beta_5\text{YOLO}_{2,t-1}+\epsilon_t.
\end{align}
$V_{t-1}$ is the volatility of the previous trading day, $M_{t-1}$ is the Google Trends activity for general terms linked to the GameStop short squeeze (\cite{lyocsa2022yolo}), and $\text{VIX}_{t-1}$ is the Chicago Board Options Exchange's (CBOE) Volatility Index. $\textbf{M1}$ is the baseline model without any Reddit or Twitter data. The models \textbf{M2}, \textbf{M3}, \& \textbf{M4} use all the parameters of \textbf{M1}, hence the $\textbf{M1}_{t-1}$ term. In \cref{sec:results}, we will extend these models with our parameters. $\textbf{M1-4}$ will be used for comparison.

\section{Results}
\label{sec:results}
\subsection{MOON parameter}
We introduce two MOON parameters. The first MOON parameter $\text{MOON}_{1,t}$, doesn't use the sentiment. It is intended to correspond to the general activity on Twitter for a given day. For each day, we collect the tweet count for our queries for a given day in $N_{t}$. For all non-trading days, we averaged the tweet count with the previous trading day. In contrast to the sentiment analysis, we included tweets with video, images, and other media attachments. As we will look at the logarithmic volatility, we took the logarithm of this quantity. To summarize, we write $\text{MOON}_{1,t}$ as:

\begin{equation}
    \text{MOON}_{1,t} = \log N_t
\end{equation}

The second parameter $\text{MOON}_{2,t}$, includes the VADER compound sentiment value $S(T)$. Here, $T$ stands for a given tweet. $S(T)=1$ if the tweet $T$ has positive sentiment, $S(T)=0$ for a neutral tweet, and $S(T)=-1$ for a negative tweet. We will take the standard deviation to get a sense of the polarity for a given day:

\begin{equation}
    \text{MOON}_{2,t} = \text{std}\left( S(T) \right)_t
\end{equation}

We tried multiple ways of combining the sentiment, but the standard deviation is one of the simplest and yielded consistent results. Other approaches we tried were: the mean, the weighted mean by public metrics (likes, retweets, quotes, and comments), and the ratio of positive or negative tweets.

\begin{figure}[H]
    \centering
    \makebox[\textwidth][c]{\includegraphics[width=1\columnwidth]{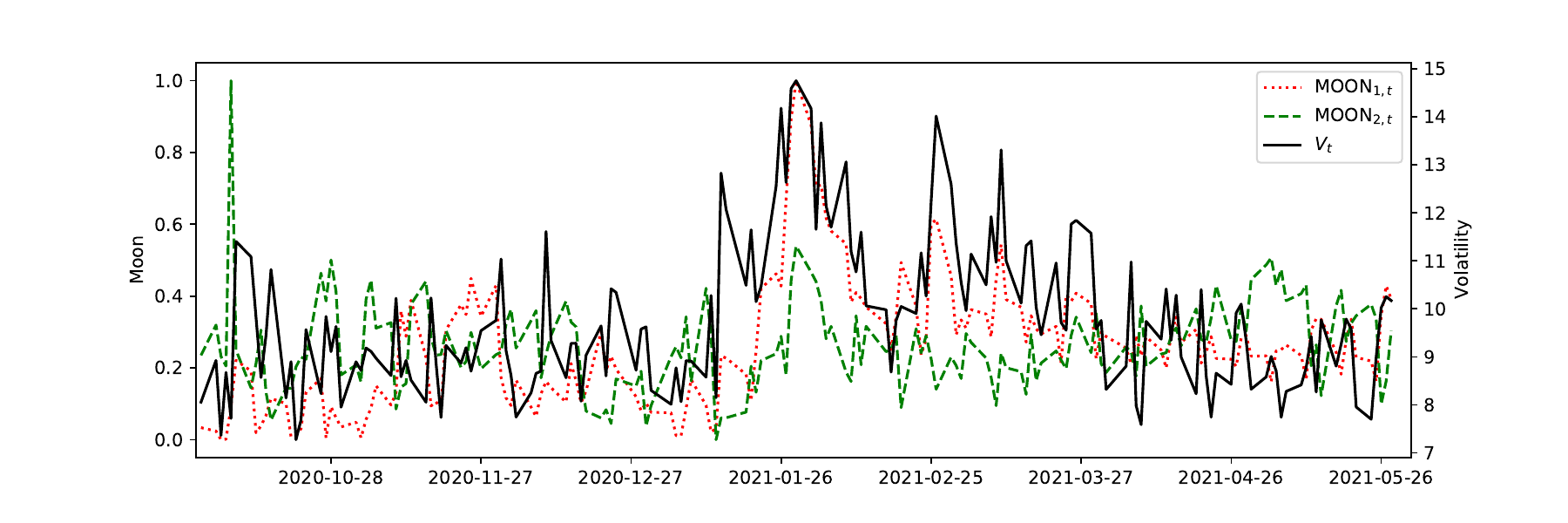}}
    \caption{Visualization of $\text{MOON}_{1,t}$, $\text{MOON}_{2,t}$, and $V_t$ for GameStop.}
    \label{fig:gme_data}
\end{figure}

\begin{figure}[H]
    \centering
    \makebox[\textwidth][c]{\includegraphics[width=1\columnwidth]{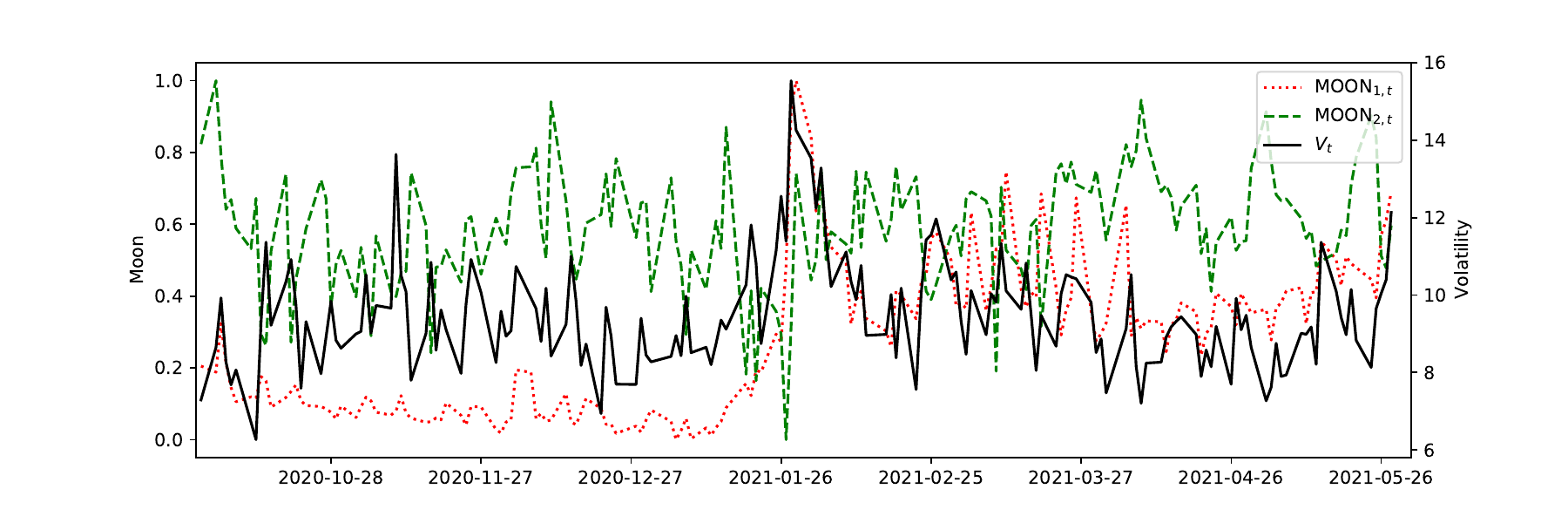}}
    \caption{Visualization of $\text{MOON}_{1,t}$, $\text{MOON}_{2,t}$, and $V_t$ for AMC.}
    \label{fig:amc_data}
\end{figure}

\begin{figure}[H]
    \centering
    \makebox[\textwidth][c]{\includegraphics[width=1\columnwidth]{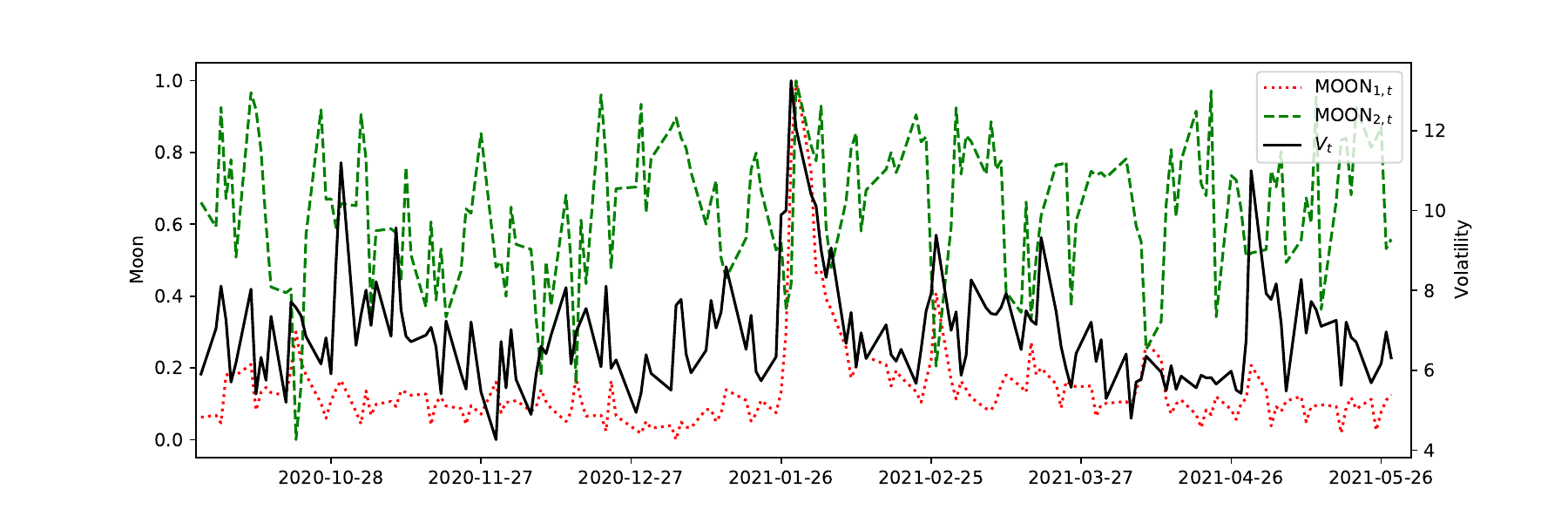}}
    \caption{Visualization of $\text{MOON}_{1,t}$, $\text{MOON}_{2,t}$, and $V_t$ for Nokia.}
    \label{fig:nok_data}
\end{figure}

\begin{figure}[H]
    \centering
    \makebox[\textwidth][c]{\includegraphics[width=1\columnwidth]{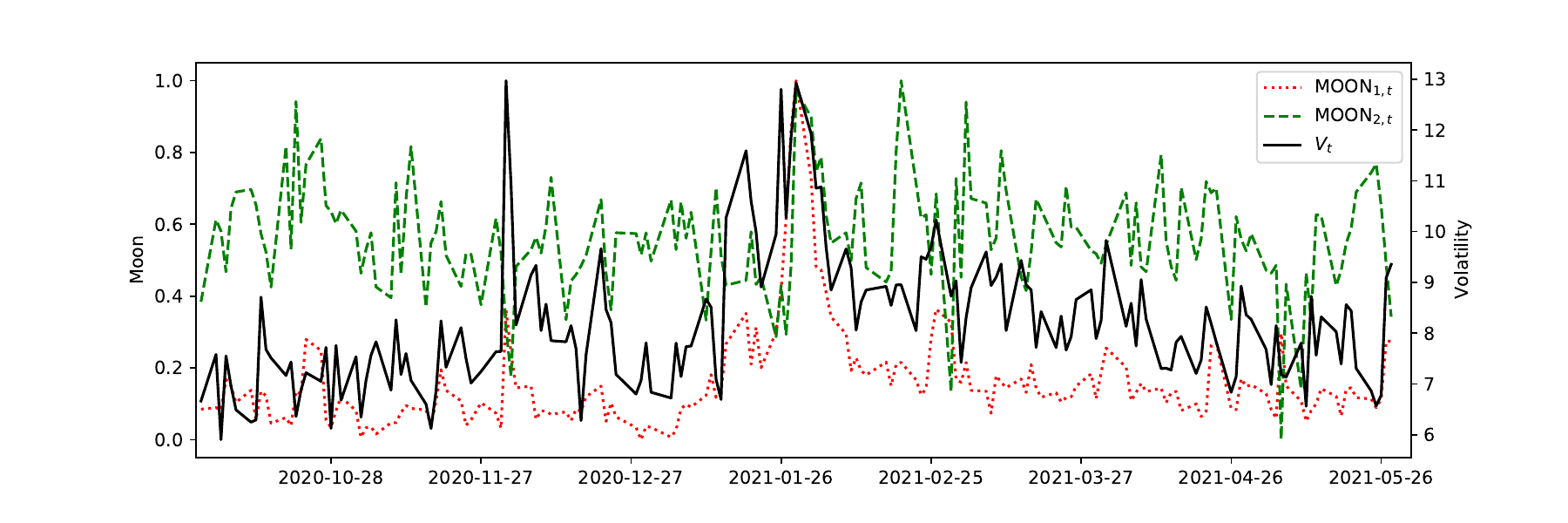}}
    \caption{Visualization of $\text{MOON}_{1,t}$, $\text{MOON}_{2,t}$, and $V_t$ for BlackBerry.}
    \label{fig:bb_data}
\end{figure}

In ~\cref{fig:gme_data,fig:amc_data,fig:bb_data,fig:nok_data}, we visualize both of our MOON variables, $\text{MOON}_{1,t}$ in red \& $\text{MOON}_{2,t}$ in blue, together with the volatility $V_t$ in black. The x-axis notes the observed time frame. While the left y-axis shows the normalized MOON variables. $\text{MOON}_{1,t}$ and $\text{MOON}_{2,t}$ were  mapped linearly to the interval $[0,1]$ to improve readability. The right y-axis shows the values of the logarithmic volatility, as described in \cref{sec:VolCal}.

The correlations in \cref{tab:GME_Corr} seem to agree with the visual representation in \cref{fig:gme_data}. There appears to be a correlation between the general Twitter activity $\text{MOON}_{1,t}$ and the volatility $V_t$, but less so for the sentiment on Twitter $\text{MOON}_{2,t}$.
This pattern continues for the three other stocks in \cref{fig:amc_data,fig:bb_data,fig:nok_data}.


\subsection{Correlations}
We investigated the Pearson correlation coefficient between $V_t$, $\text{MOON}_{1,t}$, $\text{MOON}_{2,t}$, $\text{YOLO}_{1,t}$, and $\text{YOLO}_{2,t}$. The specific values for GameStop can be seen in \cref{tab:GME_Corr}. The results for the other stocks are in the Appendix.

\begin{table}
    \centering
    \begin{tabular}{c|cccc}
        \textbf{GameStop} & $\text{MOON}_{1,t}$& $\text{MOON}_{2,t}$& $\text{YOLO}_{1,t}$& $\text{YOLO}_{2,t}$\\   
 \hline
$V_t$& 0.699 &-0.119 &0.762 &0.527 \\
$\text{MOON}_{1,t}$&  &0.088 &0.628 &0.753 \\
$\text{MOON}_{2,t}$&  & &-0.104 &0.271 \\
$\text{YOLO}_{1,t}$&  & & &0.470 \\
    \end{tabular}
    \caption{GME Correlation}
    \label{tab:GME_Corr}
\end{table}

\subsection{Regression models}
We extend the models from \cref{sec:RegMethod} by \textbf{M5}, \textbf{M6}, \textbf{M7}, and \textbf{M8}:

\begin{align}
    \textbf{M5: } V_t&=\textbf{M1}_{t-1}+\beta_6 \text{MOON}_{1,t-1}+\epsilon_t,\\
   \textbf{M6: } V_t&=\textbf{M1}_{t-1}+\beta_7 \text{MOON}_{2,t-1}+\epsilon_t,\\
    \textbf{M7: } V_t&=\textbf{M1}_{t-1}+\beta_6 \text{MOON}_{1,t-1}+\beta_7 \text{MOON}_{2,t-1}+\epsilon_t,\\
    \textbf{M8: } V_t&=\textbf{M1}_{t-1}+\beta_6 \text{MOON}_{1,t-1}+\beta_7 \text{MOON}_{2,t-1}\nonumber \\
    & \quad \quad \quad \quad \,+ \beta_8 \text{YOLO}_{1,t-1} \; +\beta_9 \text{YOLO}_{2,t-1} \; +\epsilon_t.
\end{align}

The quality of the models, as expressed through the $R^2$ value, can be seen in \cref{tab:lin_reg_R2}. The models M1-4 use Reddit data, and M5-7 use Twitter data. M8 always performs the best because it uses all variables. 
Model M6 ($\text{MOON}_{2,t}$) always performs worse than M5 ($\text{MOON}_{1,t}$), and model M4 (Reddit data) outperforms M7 (Twitter data). A notable exception is GME, where both models' $R^2$ values are similar.

\bgroup
\begin{table}[H]
\centering
\begin{tabular}{ccccccccc}
\hline
      & \multicolumn{7}{l}{GME}                               \\ \hline
      & M1    & M2    & M3    & M4    & M5    & M6    & M7    & M8    \\ \cline{2-9} 
$R^2$ value & 0.429 & 0.499 & 0.463 & 0.520 & 0.512 & 0.435 & 0.512 & 0.560 \\ \hline
\hline
      & \multicolumn{7}{l}{AMC}                               \\ \hline
      & M1    & M2    & M3    & M4    & M5    & M6    & M7    & M8    \\ \cline{2-9}  
$R^2$ value & 0.335 & 0.459 & 0.373 & 0.460 & 0.366 & 0.343 & 0.384 & 0.468 \\ \hline
\hline
      & \multicolumn{7}{l}{NOK}                               \\ \hline
      & M1    & M2    & M3    & M4    & M5    & M6    & M7    & M8    \\ \cline{2-9}  
$R^2$ value & 0.304 & 0.451 & 0.347 & 0.463 & 0.335 & 0.319 & 0.346 & 0.479 \\ \hline
\hline
      & \multicolumn{7}{l}{BB}                               \\ \hline
      & M1    & M2    & M3    & M4    & M5    & M6    & M7    & M8    \\ \cline{2-9}  
$R^2$ value & 0.453 & 0.529 & 0.481 & 0.544 & 0.492 & 0.456 & 0.499 & 0.547 \\ \hline
\end{tabular}
\caption{Linear Regression Models}
\label{tab:lin_reg_R2}
\end{table}
\egroup

\section{Discussion}
In our study, we examined the relationship between Twitter activity, the sentiment of this activity, and the price variation of the stocks of several companies, namely GameStop, AMC, BlackBerry, and Nokia, during time frames of special social media attention. 
Thereby, we have indeed found correlations between sentiment and volatility. However, a rise in Twitter activity seems more correlated to price variation. 
The sentiment parameter $\text{MOON}_{1,t}$ can still improve the overall model performance indicating that the sentiment also has insight into the volatility prediction. While a lot of this price variation is linked to general stock market factors, our findings conclude that Twitter activity can motivate more people to participate in the event and become retail investors in these stocks. 

Our findings extend the current body of knowledge on YOLO Events by making a useful comparison with the findings of \cite{lyocsa2022yolo} and extending it to Twitter. 
The study and methodology deliberately closely resemble theirs to allow for comparison. 
We found that while the more significant part of the price variation depends on other variables, increased attention on Twitter seems to impact it.
When comparing our results, we see that overall, Twitter activity is a worse predictor of stock volatility than Reddit activity.
Notably, GME had the best performance of all the examined stocks and approached Reddit data in predictability.
Even though the event originated from Reddit, Twitter was almost as good a predictor in the most attention-grabbing of the stocks: GME. For the others, Twitter performed notably worse. There seems to be a hurdle of public attention for Twitter to become a predictive factor. Otherwise, the importance of Reddit as the origin of these events remains.
As seen in \cref{tab:GME_Corr}, there is also a high correlation between Twitter activity $\text{MOON}_{1,t}$, and the volatility $V_t$ compared to the correlation between the sentiment $\text{MOON}_{2,t}$, and the volatility $V_t$ for GME. In this case, the correlation with Twitter activity $\text{MOON}_{1,t}$ is also higher than the factor corresponding to Google activity $\text{YOLO}_{2,t}$.

Brokers were in an especially awkward position during the course of these events. They are the main gateway to facilitate the trade of retail investors. Since facilitating a transaction usually takes about two business days, they were occasionally forced to stop and restrict these trades to continue fulfilling their legal security obligations as brokers.
These restrictions greatly impact the stocks and their customers (\cite{jones2021brokerages}). The online broker \enquote{Robinhood Markets, Inc.} played a special role that was especially popular among retail investors. On 28 January 2021, they restricted the trading of stocks with high volatility, including the four stocks examined by \cite{kelleher2022securities}. The role that Robinhood played during these events was investigated by \cite{pasztor2021robinhood}, \cite{tan2021democratizing}, and \cite{barber2022attention}.
Our findings show that by observing Twitter activity and sentiment, these brokers can't prepare for these occasions as well as if they observed Reddit activity directly. Only for the most-attention grabbing events, like the GameStop short squeeze, can brokers utilize Twitter to anticipate these kinds of events to better prepare for these occasions and avoid implementing trading restrictions.  
On another note, brokers can also gain a competitive advantage by offering their customers this real-time analysis of stocks. Sentiment analysis is already recognized as beneficial for stock prediction (\cite{ho2019harnessing}). Our findings can be used to signal the onset and occurrence of such special equity rallies. For stock traders, the knowledge that more extensive attention on Twitter during these events can translate to increased predictability of the volatility of the stock is also valuable when participating in these events. The finding that sentiment has a negligible impact on the predictability of price variation also opens up additional questions regarding how it can be instrumentalized during these events for stock traders.

For market regulators, these kinds of events pose an extraordinary challenge. The question of whether and how to regulate these events and manage their impact remains to be resolved and requires a careful balance between protection and freedom (\cite{campbell2016restoring}). But the fact that these events can have such an extraordinary impact on the stock market means that there has to be careful consideration about the way they are managed. Social media analysis can be useful for diagnosing whether such an event is taking place. 
If they generally want to foresee these events, we find that Reddit is the better place to pay attention. But if the goal is to spot more considerable market disruptions to react to them, we find that Twitter can be almost as good at signaling them. In combination, the two sources can even lead to better predictability in general. For market regulators that want to anticipate major events to react to them, we find that a combined observation of both Reddit and Twitter has the best results. 

In recent years, the number of bots on Twitter has increased \cite{chu2010tweeting}, so the question naturally arises: Can bots be used to manipulate the overall sentiment on Twitter for a given stock? This was also investigated by \cite{fan2020social} and \cite{cresci2019cashtag}. Therefore, the detection and filtering of bot messages could improve the proposed method here (\cite{alothali2018detecting,chavoshi2016debot}).
Especially since Twitter, one of the most influential social media platforms was acquired and privatized, leaving the potential for the instrumentalization of these platforms for stock market manipulation. Generating such an event artificially could have a similar impact to what we showed, and therefore this avenue of research should be explored. Can such an event be manufactured by, e.g., bots and still have a similar impact, or are the effects caused entirely by the actual retail investors?


This study has some potential limitations that should be considered in terms of the validity of the results.
We based our analysis on tweets collected via the Twitter API.
Due to our dependence on this API and its black-box nature, we may have introduced a selection bias. By choosing a different query, we may obtain other data and, therefore, different results.
Additionally, since our tweet selection was rudimentary, it stands to reason that the actual sentiment regarding a stock, when including memes, videos, and other media, can be different. 
We chose this tweet selection because we only focused on the sentiment of text data. An interesting question for future research is whether the value of observed sentiment can be substantially increased through a better sentiment calculation process, e.g., finding a way to meaningfully include memes and pictures or a better Tweet selection process. Future research may try to solve these problems by having approaches to include memes and other media in this analysis, as done by \cite{french2017image}.
The change in observed sentiment could alter our analyzed connections between our parameters and stock price volatility, which could lead to different conclusions from this research.

\section{Conclusion}
In recent years we observed several occasions where certain stocks seem to be strongly influenced by coordinated movements on social media, called YOLO-Events.
This study investigates the link between stock volatility, Twitter activity, and sentiment regarding stocks during YOLO-Events. We do this by constructing parameters from the observed sentiment and activity and correlating them to approximated stock volatility. Additionally, we created linear regression models with these parameters.

Regarding our research question, we find that the sentiment doesn't improve the performance of these models significantly most of the time. During the biggest of the events, sentiment didn't improve the model performance at all. Generally, the activity on Twitter and stock volatility have a high correlation, but we also find that Reddit is the better predictor for these events.
Our findings point to an existing link between especially Twitter activity and stock volatility and a weak link to Twitter sentiment.

\section{Acknowledgements}
We thank the authors of \cite{lyocsa2022yolo} for corespondents and for making their data available to us. We thank the Digital and Data Literacy in Teaching Lab and the Stiftung Innovation in der Hochschullehre for their expertise and financial support.

\newpage
\section{Appendix}
\label{sec:appendix}

\begin{table}[h]
    \centering
    \begin{tabular}{c|cccc}
         \textbf{AMC} & $\text{MOON}_{1,t}$& $\text{MOON}_{2,t}$& $\text{YOLO}_{1,t}$& $\text{YOLO}_{2,t}$\\   
 \hline
$V_t$& 0.459 &-0.363 &0.618 &0.417 \\
$\text{MOON}_{1,t}$&  &0.057 &0.593 &0.688 \\
$\text{MOON}_{2,t}$&  & &-0.257 &0.068 \\
$\text{YOLO}_{1,t}$&  & & &0.574 
    \end{tabular}
    \caption{AMC  Correlation}
    \label{tab:my_label}
\end{table}

\begin{table}[h]
    \centering
    \begin{tabular}{c|cccc}
        \textbf{NOK} & $\text{MOON}_{1,t}$& $\text{MOON}_{2,t}$& $\text{YOLO}_{1,t}$& $\text{YOLO}_{2,t}$\\   
 \hline
$V_t$& 0.608 &-0.061 &0.509 &0.391 \\
$\text{MOON}_{1,t}$&  &-0.118 &0.464 &0.779 \\
$\text{MOON}_{2,t}$&  & &-0.102 &0.137 \\
$\text{YOLO}_{1,t}$&  & & &0.401 
    \end{tabular}
    \caption{NOK  Correlation}
    \label{tab:my_label}
\end{table}

\begin{table}[h]
    \centering
    \begin{tabular}{c|cccc}
         \textbf{BB} & $\text{MOON}_{1,t}$& $\text{MOON}_{2,t}$& $\text{YOLO}_{1,t}$& $\text{YOLO}_{2,t}$\\   
 \hline
$V_t$& 0.740 &-0.044 &0.658 &0.517 \\
$\text{MOON}_{1,t}$&  &0.101 &0.691 &0.778 \\
$\text{MOON}_{2,t}$&  & &-0.109 &0.187 \\
$\text{YOLO}_{1,t}$&  & & &0.439 
    \end{tabular}
    \caption{BB  Correlation}
    \label{tab:my_label}
\end{table}

\newpage

\bibliographystyle{agsm}
\bibliography{main}

\end{document}